\documentclass[pdflatex,sn-mathphys-num]{sn-jnl}

\usepackage{graphicx}%
\usepackage{multirow}%
\usepackage{amsmath,amssymb,amsfonts}%
\usepackage{amsthm}%
\usepackage{mathrsfs}%
\usepackage[title]{appendix}%
\usepackage{xcolor}%
\usepackage{textcomp}%
\usepackage{manyfoot}%
\usepackage{booktabs}%
\usepackage{algorithm}%
\usepackage{algorithmicx}%
\usepackage{algpseudocode}%
\usepackage{listings}%
\UseRawInputEncoding
\usepackage{float}%
\usepackage{subcaption}

\theoremstyle{thmstyleone}%
\theoremstyle{thmstyletwo}%

\theoremstyle{thmstylethree}%

\usepackage{url}

\begin{document}

\title[Article Title]{Multi-Attention Stacked Ensemble for Lung Cancer Detection in CT Scans}


\author*[1]{\fnm{Uzzal} \sur{Saha}}\email{mt2302101017@alum.iiti.ac.in}

\author[1]{\fnm{Surya} \sur{Prakash}}\email{surya@iiti.ac.in}


\affil[1]{\orgdiv{Department of Computer Science and Engineering}, \orgname{ Indian Institute of Technology Indore}, \orgaddress{\street{Simrol}, \city{Indore}, \postcode{453552}, \state{Madhya Pradesh}, \country{India}}}




\abstract{In this work, we address the challenge of binary lung nodule classification (benign vs. malignant) using CT images by proposing a multi‐level attention stacked ensemble of deep neural networks. Three pretrained backbones: EfficientNet V2 S, MobileViT XXS, and DenseNet201 are each adapted with a custom classification head tailored to 96 x 96 pixel inputs. A two‐stage attention mechanism learns both model‐wise and class‐wise importance scores from concatenated logits, and a lightweight meta‐learner refines the final prediction. To mitigate class imbalance and improve generalization, we employ Dynamic Focal Loss with empirically calculated class weights, MixUp augmentation during training, and Test‐Time Augmentation at inference. Experiments on the LIDC-IDRI dataset demonstrate exceptional performance, achieving 98.09\% accuracy and 0.9961 AUC, representing a 35\% reduction in error rate compared to state-of-the-art methods. The model exhibits balanced performance across sensitivity which is 98.73\% and specificity which is 98.96\% with particularly strong results on challenging cases where radiologist disagreement was high. Statistical significance testing confirms the robustness of these improvements across multiple experimental runs. Our approach can serve as a robust, automated aid for radiologists in lung cancer screening.}




\maketitle

\section{Introduction}\label{sec1}

Lung cancer remains the leading cause of cancer‐related mortality worldwide, yet early detection via low‐dose CT screening can drastically improve patient outcomes \cite{siegel2023cancer}. Within CT images, pulmonary nodules exhibit a wide range of shapes, textures, and intensities, making it difficult to distinguish benign from malignant lesions. Moreover, publicly available datasets such as LIDC-IDRI are heavily skewed toward benign cases, and variations in scanner protocols and slice thickness further aggravate heterogeneity \cite{armato2011lung}. Traditional machine‐learning pipelines relying on handcrafted features struggle to generalize across these variations, while single‐backbone convolutional neural networks though powerful—often overfit to dominant patterns and fail to capture complementary characteristics of subtle nodules.

Simple ensemble techniques (e.g., majority voting or uniform averaging of logits) attempt to combine multiple CNN predictions but treat each model’s output equally, ignoring that certain architectures may excel at recognizing specific visual attributes (e.g., fine texture vs. global shape). As a result, existing approaches typically plateau around mid-90s in overall accuracy, often at the cost of lower sensitivity for malignant nodules. To overcome these limitations, we propose a Multi-Attention Stacked Ensemble (MASE) that dynamically weights both entire models and individual class predictions. Specifically, three state-of-the-art backbones—DenseNet-201, EfficientNet V2 S, and MobileViT XXS are each adapted with a lightweight classification head optimized for 96 × 96 CT patches. During inference, concatenated logits from all three are first passed through a “model-level” attention layer, which learns to emphasize more reliable network outputs; a subsequent “class-level” attention layer highlights discriminative features for benign versus malignant classes. A small meta-learner then fuses these attended representations into a final prediction.

To further improve robustness against class imbalance and overfitting, the MASE framework integrates Dynamic Focal Loss (with empirically computed class weights), MixUp augmentation during training, and Test-Time Augmentation at inference. In extensive experiments on LIDC-IDRI, MASE consistently surpasses each standalone backbone and uniform ensembling, achieving higher sensitivity for malignant nodules while maintaining strong specificity. These results suggest that dynamically attending to model and class cues can significantly bolster automated lung cancer screening systems, offering a reliable second-reader tool to reduce inter-observer variability.

The remainder of this paper is organized as follows. Section 2 reviews related work; Section 3 describes the dataset and preprocessing; Section 4 details model architectures and loss functions; Section 5 outlines training and evaluation protocols; Section 6 presents experimental results and analysis; Section 7 discusses findings and limitations; and Section 8 concludes.

\section{Related work}\label{sec2}

Early computer‐aided diagnosis (CAD) pipelines for lung nodules followed a multi‐stage workflow: lung segmentation (often via Hounsfield‐unit thresholding and morphological operations), candidate detection (gray‐level thresholding, shape analysis), handcrafted feature extraction, and classical classification (SVM, Random Forest, k‐NN) \cite{el2013computer, kostis2013relationships}. Handcrafted descriptors included intensity statistics (mean, variance, skewness), shape attributes (sphericity, elongation, perimeter‐to‐area ratio), and texture features such as Gray Level Co‐occurrence Matrix (GLCM) and Local Binary Patterns (LBP) \cite{Way2012texture, Zhao2010wavelet, Murphy2009shape}. Although some hybrid approaches e.g., Froz et al.’s artificial crawler and rose‐diagram method reached up to 94.4 \% accuracy on small cohorts \cite{froz2017lung}, these methods failed to generalize to large, heterogeneous datasets like LIDC‐IDRI, due to scanner variability and limited capacity to learn complex visual patterns.

The emergence of convolutional neural networks (CNNs) shifted focus to end‐to‐end learning from raw CT data. Early 2D CNN models applied patch‐based classification on axial slices. Shen et al.\ \cite{Shen2015multi} achieved ~90 \% accuracy on a subset of LIDC‐IDRI using a multi‐scale 2D CNN. Subsequent work fine‐tuned deeper architectures (ResNet‐50, DenseNet‐121, Inception‐V3) pretrained on ImageNet, reporting 92 \%–94 \% accuracy with focal loss or weighted cross‐entropy \cite{Nam2019deep}. Three‐dimensional CNNs (e.g., 32³‐voxel cubes) further improved context modeling: Liao et al.\ \cite{Liao2020automatic} combined a 3D ResNet backbone with a recurrent network to achieve 94 \% accuracy on LIDC‐IDRI. More recently, transformer‐inspired models (e.g., Vision Transformer variants) have been applied to volumetric patches; Li et al.\ \cite{Li2022translung} reported 95 \% accuracy by integrating self‐attention over 3D nodule volumes. Despite these strides, single‐backbone approaches frequently plateau in the mid‐90s, especially under severe class imbalance.

Ensembling multiple deep networks via averaging logits or majority voting has become a common strategy to boost classification robustness. Sagi et al.\ \cite{Sagi2018ensemble} combined three distinct 2D CNNs on LIDC‐IDRI patches, reporting a 95 \% overall accuracy. Antolović et al.\ \cite{Antolovic2019attention} introduced a gated ensemble where a trainable weighting module assigns per input model importance, yielding a 1–2 \% improvement on lung nodule cohorts. Zhang et al.\ \cite{Zhang2021multi} extended this by applying a single‐level attention mechanism over concatenated logits, achieving 96 \% accuracy. However, most existing ensembles (even attention‐based) treat each model’s output as a monolithic vector neglecting the possibility that certain models may be more reliable for specific classes. Very few studies incorporate both model‐level and class‐level attention within a unified framework.

Attention modules originally popularized by transformers allows network to focus on diagnostically relevant regions or feature channels. In lung nodule analysis, attention gates have highlighted subtle nodule boundaries without explicit supervision \cite{li2019selective, schlemper2018attention}. Focal Loss \cite{Lin2017focal} mitigates class imbalance by down weighting easy examples, and Dynamic Focal Loss \cite{Zhang2021dynamicfocal} further adjusts the $\alpha$ parameter per class during training. MixUp augmentation \cite{Zhang2018mixup} creates interpolated samples that improve generalization. Test‐Time Augmentation (TTA) \cite{Wang2018tta} aggregates predictions over random flips and rotations, reducing inference variance; Chen et al.\ \cite{Chen2020tta} reported a 1 \% AUC improvement with TTA. While these techniques have become widespread, few studies combine attention‐based ensembling with dynamic loss and augmentation strategies in a single pipeline for lung nodule classification.

Traditional methods reliant on handcrafted features struggle to generalize across diverse CT protocols, and single CNNs often plateau due to class imbalance and limited feature complementarity. Simple ensembles and single‐level attention ensembles improve robustness modestly but neglect class‐specific reliability. Our Multi‐Attention Stacked Ensemble addresses these gaps by integrating three state‐of‐the‐art backbones (DenseNet‐201, EfficientNet V2 S, MobileViT XXS) via customized adapters, applying attention at both model and class levels for richer fusion, and leveraging Dynamic Focal Loss, MixUp, and TTA to combat imbalance and enhance generalization. This holistic framework outperforms prior approaches on LIDC‐IDRI, demonstrating improved sensitivity for malignant nodules while maintaining high specificity.

\section{Dataset and Preprocessing}

The Lung Image Database Consortium and Image Database Resource Initiative (LIDC‐IDRI) is a publicly available repository of thoracic CT scans annotated by four expert radiologists, and is widely used for developing and evaluating lung nodule classification methods \cite{armato2011lung}.

\begin{figure}[htbp]
    \centering
    \includegraphics[width=0.9\linewidth]{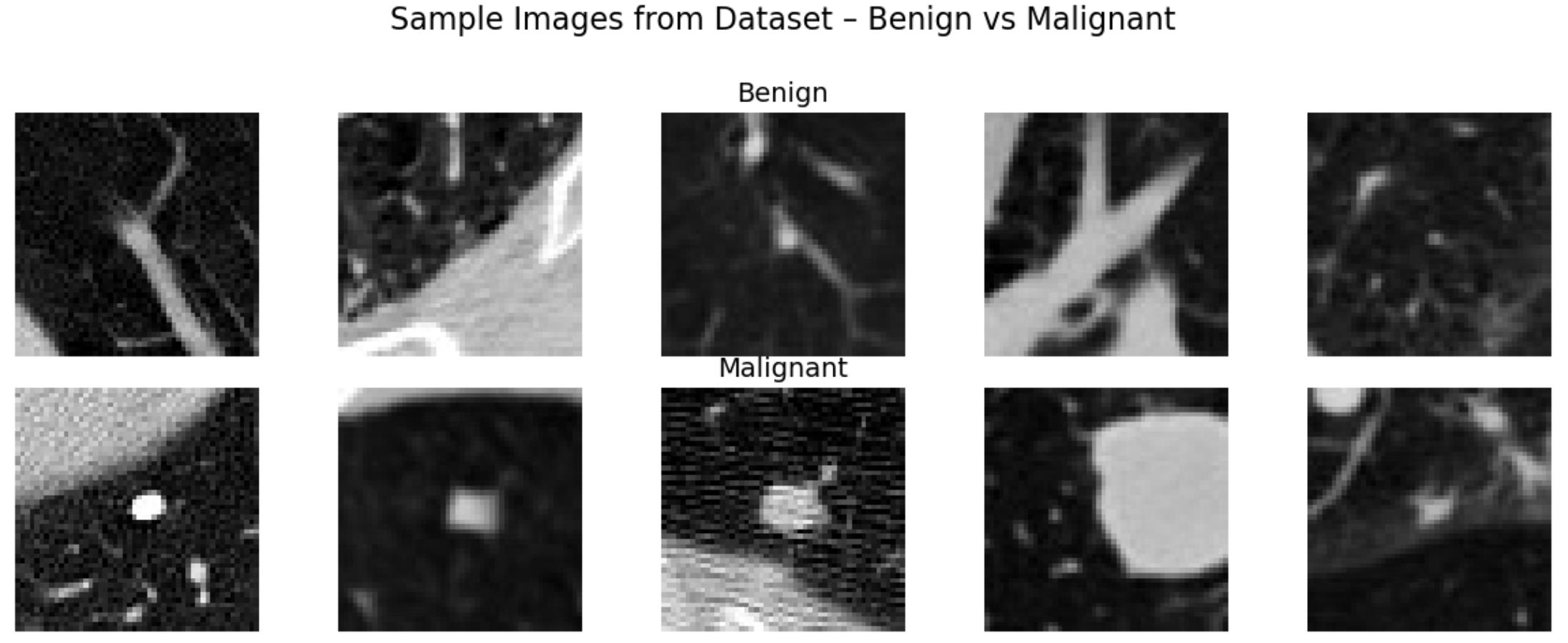}
    \caption{Sample CT image patches extracted from the LIDC-IDRI dataset, illustrating benign (top row) and malignant (bottom row) lung nodules. These examples demonstrate the subtle visual differences that exist between the two classes and emphasize the difficulty of the classification task.}
    \label{fig:dataset_visual}
\end{figure}

\begin{figure}[htbp]
    \centering
    \begin{subfigure}[t]{0.45\linewidth}
        \centering
        \includegraphics[width=\linewidth]{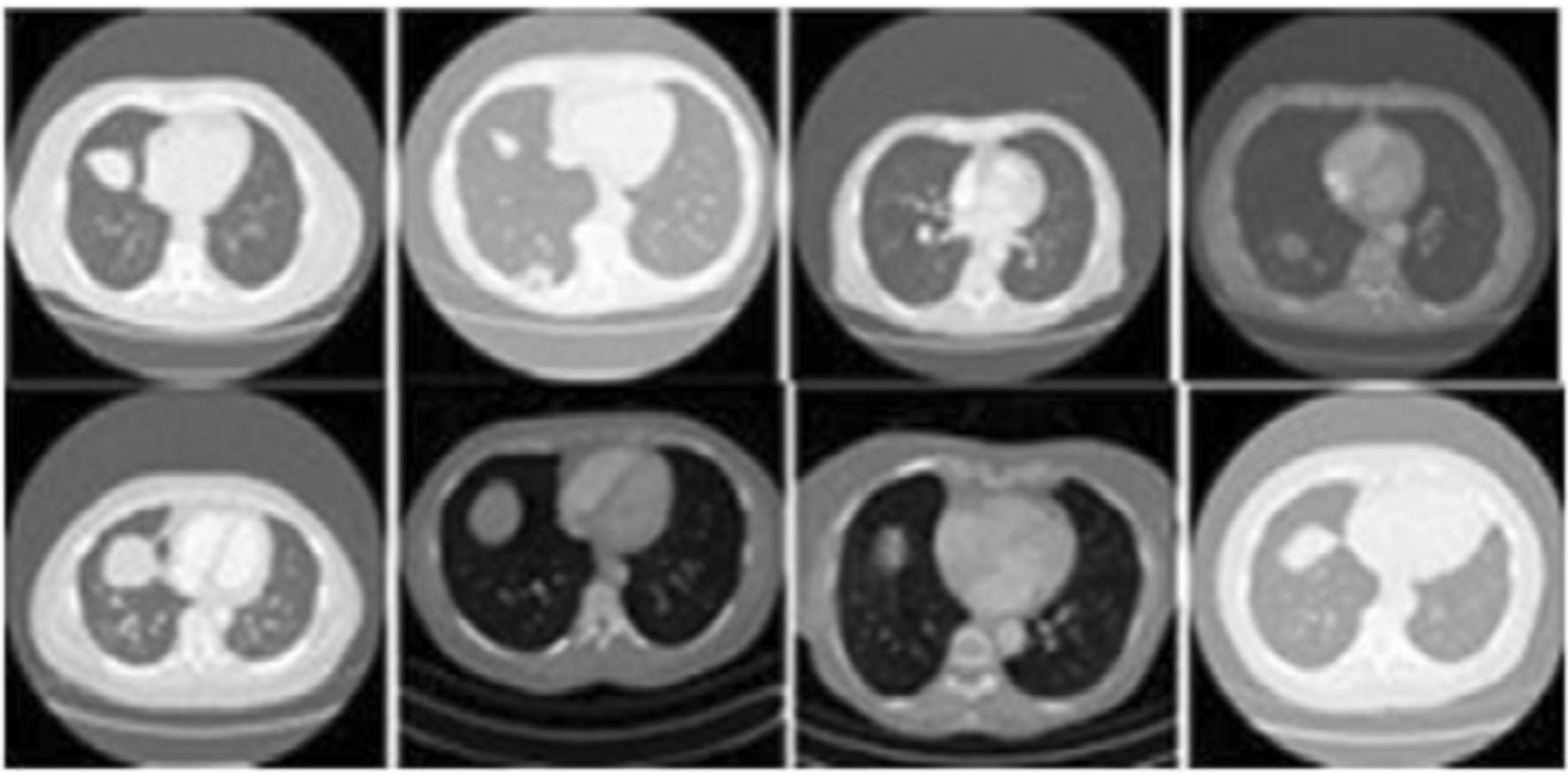}
        \caption{Benign lung CT scan images from the LIDC-IDRI dataset.}
        \label{fig:benign_ct}
    \end{subfigure}
    \hfill
    \begin{subfigure}[t]{0.45\linewidth}
        \centering
        \includegraphics[width=\linewidth]{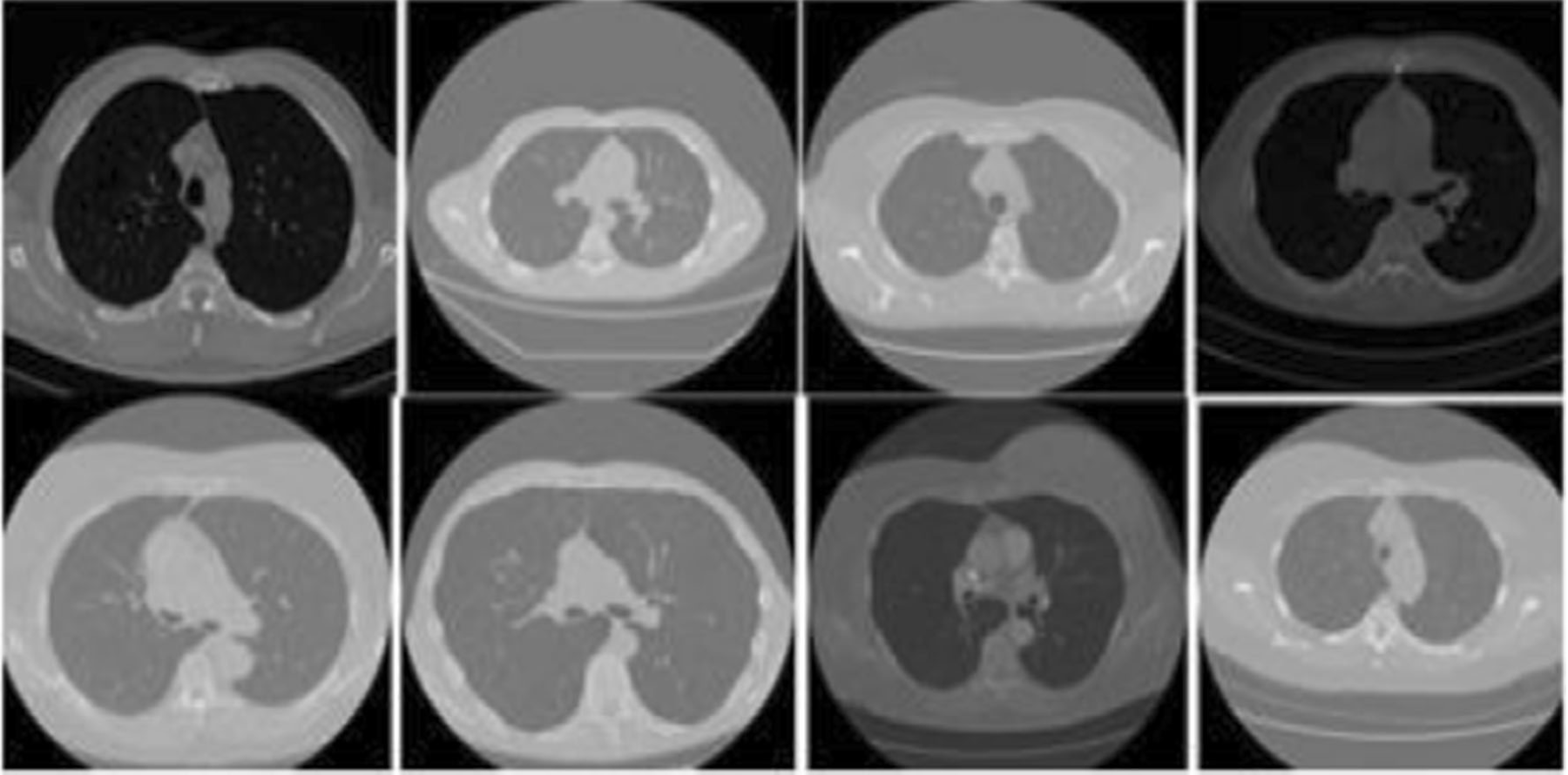}
        \caption{Malignant lung CT scan images from the LIDC-IDRI dataset.}
        \label{fig:malignant_ct}
    \end{subfigure}
    \caption{Representative axial CT slices for benign and malignant cases. }
    \label{fig:ct_scan_comparison}
\end{figure}

From the 1018 CT scans collected from 1010 patients. Nodules were binarized by averaging the radiologists’ malignancy scores ranging from 1 to 5 and thresholding at 3. Our patient‐level stratified split yields 5187 training samples among them 4342 were benign and 845 malignant, we have total 1297 validation samples containing 1073 benign and 224 malignant cases, and 1622 test samples containing 1340 benign and 282 malignant cases, maintaining an approximate 64:16:20 ratio. This division ensures sufficient data for model training while providing robust validation and test sets for performance evaluation.The dataset is organized into two classes: benign and malignant nodules. As shown in Figure~\ref{fig:dataset_visual}, a clear visual class imbalance and structural diversity are evident between the two categories. In addition to nodule patches, Figure~\ref{fig:ct_scan_comparison} presents full CT scan slices for benign and malignant cases, providing broader anatomical context for the classification task.

All patches were resized to \(96\times96\) pixels using bilinear interpolation to preserve subtle texture details while balancing GPU memory constraints. Following Inspired by prior work on transferring knowledge from models pretrained on natural images \cite{raghu2019transfusion}, we normalized the CT scan intensities using the standard ImageNet statistics. Specifically, we used a mean of [0.485, 0.456, 0.406] and a standard deviation of [0.229, 0.224, 0.225]. To improve generalization and reduce overfitting, we applied a diverse set of on-the-fly data augmentations during training. These included geometric transformations such as random horizontal and vertical flips, rotations up to ±30 degrees, along with affine transformations that include translations of up to 10\%, scaling factors ranging from 0.9 to 1.1, and shear angles within ±5 degrees. In addition, we applied photometric augmentations using color jittering, which randomly adjusted the brightness, contrast, and saturation of the images by up to ±0.2, and modified the hue within a range of ±0.05. To improve robustness against occlusions and irregularities in the data, we incorporated Random Erasing with a probability of 0.2, where small random regions covering between 2\% and 20\% of the image area were masked out. To enhance generalization, we adopted the MixUp data augmentation strategy proposed by Zhang et al.~\cite{zhang2017mixup}, which generates synthetic training examples by linearly combining pairs of images and their corresponding labels. For each minibatch during training, with a probability of 0.7, a mixing coefficient $\lambda$ is sampled from a Beta distribution with both shape parameters set to 0.4:

\begin{equation}
\lambda \sim \mathrm{Beta}(\alpha,\alpha)
\label{beta}
\end{equation}

Given two randomly selected training samples $(x_i, y_i)$ and $(x_j, y_j)$, the augmented sample $(\tilde{x}, \tilde{y})$ is computed as:
\begin{equation}
\tilde{x} = \lambda \cdot x_i + (1-\lambda) \cdot x_j
\label{e2}
\end{equation}

\begin{equation}
\tilde{y} = \lambda \cdot y_i + (1-\lambda) \cdot y_j
\label{e3}
\end{equation}

As defined in Equations~\ref{e2} and~\ref{e3}, this approach encourages the model to behave linearly between training examples, which is particularly beneficial in medical imaging tasks where class boundaries—such as between benign and malignant nodules—can be ambiguous.

During validation and testing, only resizing and normalization were performed to maintain consistency and prevent any data augmentation from influencing the evaluation results. To address the $\sim$5:1 benign–malignant imbalance, we used a WeightedRandomSampler during training. Class weights were computed as:
\begin{equation}
w_c = \frac{1 / n_c}{\sum_{k} (1 / n_k)} \times C,
\label{e4}
\end{equation}
yielding weights [0.3258, 1.6742] for benign and malignant classes. Equation~\ref{e4} shows that this sampler oversamples malignant cases each epoch, mitigating majority class bias more effectively than simple loss weighting or undersampling.

\section{Proposed Method}

The proposed Multi-Attention Stacked Ensemble (MASE) architecture integrates a total of three state-of-the-art convolutional neural networks as its foundation. Each base model was selected for its complementary strengths in feature extraction and representation learning, providing diverse perspectives on the lung nodule classification task. 

\begin{figure}[htbp]
    \centering
    \includegraphics[width=\linewidth, trim=0cm 10cm 0cm 5cm, clip]{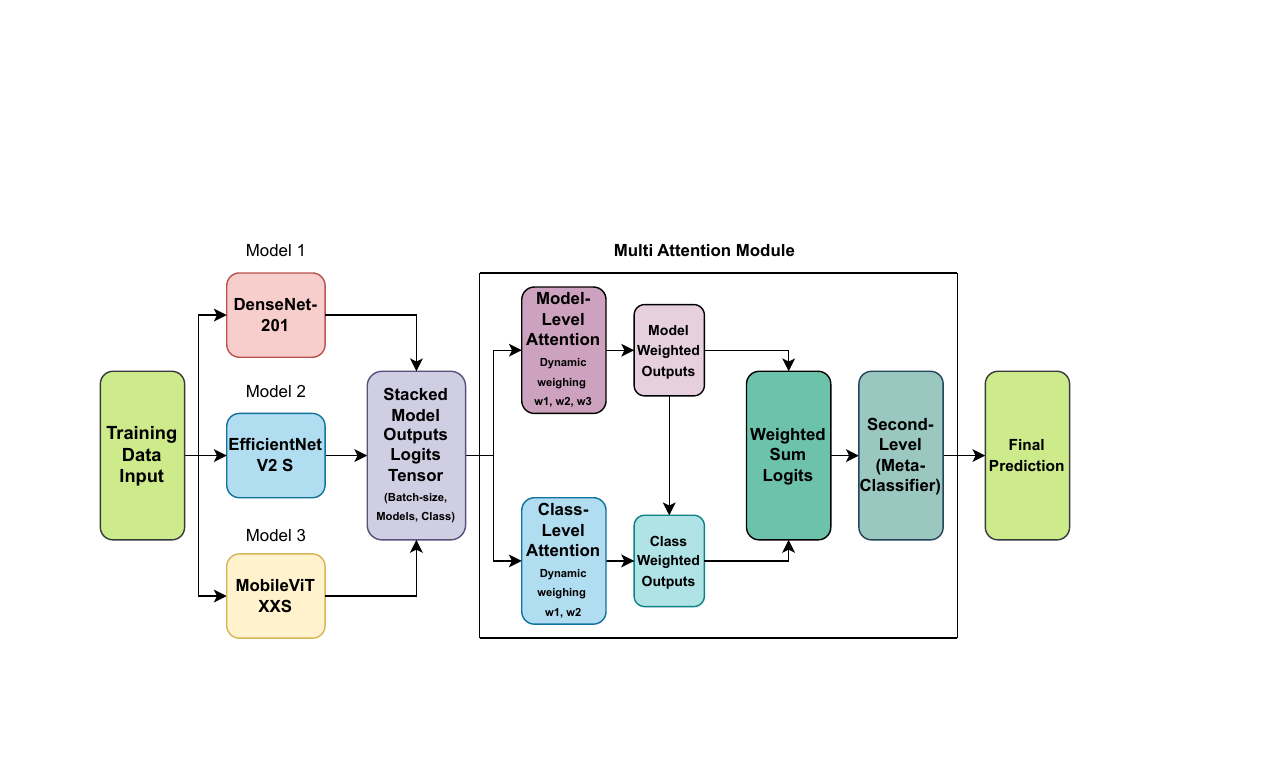}
    \caption{End-to-end pipeline of the proposed Multi-Attention Stacked Ensemble (MASE) framework for lung nodule classification. The architecture begins with training data processed by three distinct CNN backbones: DenseNet-201, EfficientNetV2-S, and MobileViT-XXS. Their outputs are fused using multi attention mechanisms (model-level and class-level attention), followed by a meta-classifier that generates the final prediction.}
    \label{fig:mase_pipeline}
\end{figure}

An overview of the complete classification pipeline is shown in Figure~\ref{fig:mase_pipeline}.

DenseNet-201 is employed as one of the backbone networks in our ensemble framework due to its ability to encourage feature reuse and alleviate the vanishing gradient problem through dense connectivity~\cite{huang2017densely}. The architecture comprises four dense blocks with 6, 12, 48, and 32 layers, respectively, interleaved with transition layers that perform downsampling. Each layer within a dense block receives the concatenated outputs of all preceding layers, resulting in improved gradient flow and more compact parameter usage compared to traditional CNNs. To adapt DenseNet-201 for lung nodule classification, we modified the original architecture in several ways. The default classification head was replaced with a custom binary classifier tailored to our task, as detailed in Section 4.2. The input resolution was adjusted to 96×96 pixels to match the size of the CT patches used in our dataset. Furthermore, we utilized ImageNet pre-trained weights for initialization, except for the final classification layer, which was randomly initialized to accommodate the binary nature of our problem. These adjustments allow DenseNet-201 to maintain its multiscale feature learning capabilities while being specialized for identifying subtle patterns in lung nodules.

EfficientNetV2-S is adopted as the second backbone in our ensemble for its balance between model accuracy and computational efficiency~\cite{tan2021efficientnetv2}. Building upon the original EfficientNet, this architecture integrates both MBConv and Fused-MBConv blocks and follows a progressive learning strategy that enhances training speed and generalization. The model comprises seven stages, each containing blocks with varying expansion factors, strides, and channel widths, beginning with a standard convolutional layer and concluding in a classification head. To tailor EfficientNetV2-S for lung nodule classification, we made several architectural adjustments. The first convolutional layer’s stride was reduced from 2 to 1 while maintaining a 3×3 kernel with padding 1, allowing the network to retain finer spatial features that are critical for identifying small nodules. We also replaced the default classifier head with a task-specific binary classifier, and pretrained weights from ImageNet were used to initialize all layers except those newly added or modified. These changes enable the model to handle 96×96 CT scan inputs more effectively, preserving spatial granularity in early layers and maintaining the model’s strength in extracting relevant features for classification.

MobileViT-XXS serves as the third backbone in our ensemble, blending convolutional inductive biases with transformer‑based global attention for an efficient yet powerful feature extractor~\cite{mehta2021mobilevit}.  It begins with a MobileNet‑style stem, then alternates between local convolutions (point‑wise and depth‑wise) and lightweight transformer blocks that unfold spatial maps into token sequences and refold them after self‑attention. This design captures fine‑grained local details and long‑range context in a single model. For lung nodule classification, we replaced its default classification head with a task‐specific binary classifier, updated the input pipeline to handle 96×96 CT patches, and initialized with ImageNet weights for all unchanged layers. These adaptations allow MobileViT-XXS to preserve its global reasoning capability while concentrating on subtle nodule characteristics amid surrounding lung tissue.

A key innovation of our approach is the development of a unified model adapter that replaces the standard classification heads of each base model. This custom head was specifically designed to address key challenges in medical image classification, such as limited annotated data and overfitting, and the critical need for well‑calibrated confidence estimates. The adapter begins with a dropout layer set to 50 percent, providing strong regularization by randomly omitting half of the features. Next comes a linear projection that compresses the backbone’s features into a fixed 256‑dimensional space, this projection step ensures that all three models, even though they use different architectures, produce outputs on the same scale. We use Layer Normalization rather than BatchNorm to stabilize training when batch sizes are small, as is typical in medical imaging, and follow this with a ReLU activation to introduce non‑linearity. Before the final classification, a second dropout layer at 30 percent further discourages over‑reliance on any single feature. This two‑stage dropout strategy not only combats overfitting but also encourages the adapter to learn a more robust, distributed representation of the input. By projecting each backbone into a common feature space, we enable direct comparison and seamless fusion in the attention ensemble that follows. Moreover, the consistent 256‑dimensional bottleneck significantly reduces the number of parameters in the final layers, acting as a form of architectural regularization that supports better generalization. LayerNorm’s contribution to stable gradient flow also yields more reliable probability outputs, which is crucial in clinical settings where decision confidence must be trustworthy. Altogether, these design choices transform general‑purpose vision architectures into specialized, parameter‑efficient classifiers that are well suited to the unique demands of lung nodule detection.

\subsection{Proposed Ensemble Method}

The core innovation of our architecture lies in the Multi-Attention Stacked Ensemble, which adaptively integrates predictions from the three base models using a dual-attention mechanism. Unlike traditional ensembles with static weights or simple averaging, our method dynamically modulates each model and class contribution based on the input, enhancing decision precision.

\begin{figure*}[t]
    \centering
    \includegraphics[width=.8\textwidth]{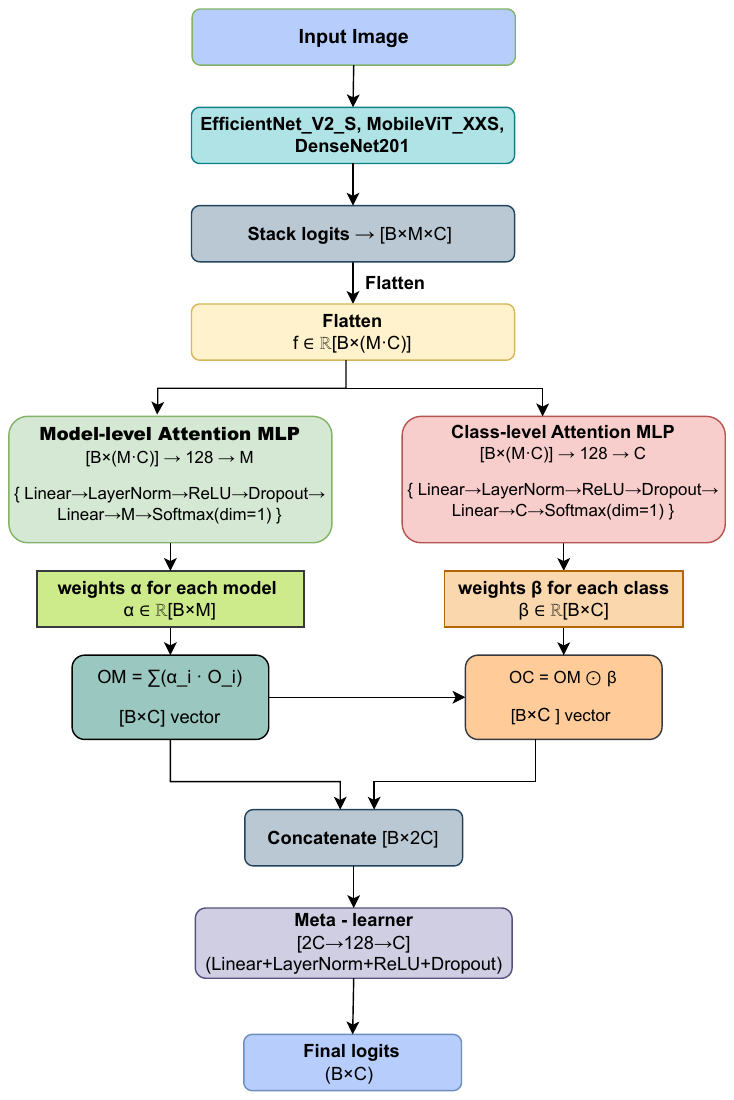}
    \caption{Schematic diagram of the dual attention mechanism in our Multi-Attention Stacked Ensemble (MASE) architecture. The model-level attention assigns weights to base models, while the class-level attention refines predictions at the class level for each input sample.}
    \label{fig:multi_attention_diagram}
\end{figure*}

The model-level attention module dynamically assigns importance weights to each base model's predictions, allowing the ensemble to leverage the unique strengths of different architectures for varying nodule characteristics. Rather than using fixed weights, this approach recognizes that certain models may excel on specific nodule subtypes for example, MobileViT-XXS's transformer components might prove more valuable for cases requiring global contextual reasoning, while DenseNet-201's dense connectivity could better capture fine-grained textural patterns.The module computes attention weights \(w_1\), \(w_2\), and \(w_3\) for DenseNet-201, EfficientNetV2-S, and MobileViT-XXS respectively, where these weights sum to 1 and represent each model's relative contribution to the final prediction. Importantly, these weights are calculated dynamically for each input sample rather than remaining fixed across the dataset.

Given the stacked outputs from three base models with tensor shape \([B, M, C]\) (where \(B\) is batch size, \(M\) is number of models, and \(C\) is number of classes), the mechanism first flattens this representation to \([B, M \times C]\). This flattened vector then passes through an MLP consisting of: linear transformation \(\mathbb{R}^{M\times C} \rightarrow \mathbb{R}^{128}\), LayerNorm, ReLU activation, dropout layer with a rate of 0.3, linear transformation \(\mathbb{R}^{128} \rightarrow \mathbb{R}^{M}\), and softmax activation. The resulting output provides attention weights \([w_1, w_2, w_3]\) for each sample. The inclusion of LayerNorm and dropout within the attention MLP promotes stable training while preventing overfitting to specific training patterns. The 128-dimensional intermediate layer offers sufficient capacity for learning complex relationships between model outputs while maintaining computational efficiency.

While model-level attention decides which backbone to trust, class-level attention provides a second weighting dimension by modulating the importance of each class prediction. This allows the ensemble to selectively emphasize certain classes based on input characteristics for instance, boosting malignant predictions in ambiguous cases to increase sensitivity, which is crucial in medical screening where false negatives are costlier than false positives.

The class-level attention uses the same flattened representation $[B,M \times C][B, M \times C]$ $[B,M \times C]$ and passes it through a parallel MLP: $\mathbb{R}^{M\times C}\rightarrow 128$, LayerNorm → ReLU → Dropout(0.3) → $\mathbb{R}^{C}$ → Softmax, producing per-class weights. These weights modulate the model-weighted predictions element-wise, introducing additional non-linearity that enables more complex decision boundaries than model-level attention alone. This dual-attention approach provides learned, input-dependent re-scaling that captures patterns beyond standard softmax operations.

The meta-learner module represents the final integration stage of our ensemble architecture, synthesizing outputs from both attention mechanisms to produce refined class predictions. Rather than directly utilizing attention-weighted outputs, we introduce a dedicated meta-learner capable of discovering complex interactions between the dual attention streams.
Our approach concatenates the model-level and class-level weighted outputs, creating a feature representation of dimensionality $2C$. 
This enriched representation undergoes processing through a compact MLP architecture: $\mathbb{R}^{2C} \rightarrow \mathbb{R}^{128}$ with LayerNorm and ReLU activation, followed by dropout regularization (rate 0.3) and a final projection to $\mathbb{R}^{C}$ for class logits. The integration process follows a principled approach where model-level attention first produces weighted combinations of base model outputs, followed by class-level modulation of these weighted predictions. Formally, given model-weighted output $\mathbf{m} = \sum_i (\mathbf{s}_i \cdot w_i^{(m)})$ and class-weighted output $\mathbf{c} = \mathbf{m} \odot \mathbf{w}^{(c)}$, the meta-learner processes the concatenated representation $[\mathbf{m}, \mathbf{c}]$ to generate final predictions.
This architecture enables the meta-learner to exploit complementary information from both attention mechanisms while maintaining the flexibility to adaptively weight their relative contributions based on input characteristics, ultimately enhancing the ensemble's discriminative capacity.

Given the inherent class imbalance in lung nodule datasets, we adopt a dynamic focal loss strategy that integrates class weighting with focal modulation to address both statistical bias and learning difficulty~\cite{lin2018focallossdenseobject}. This approach proves particularly valuable in medical imaging where minority class detection is paramount.

We derive class weights inversely proportional to class frequencies, effectively amplifying the learning signal from underrepresented samples:

\begin{equation}
\text{class\_weights}_k = \frac{1 / \text{class\_counts}_k}{\sum_j (1 / \text{class\_counts}_j)} \times \text{num\_classes}
\label{e5}
\end{equation}

These weights (0.33 for benign, 1.67 for malignant) serve as the $\alpha$ parameter in our focal loss formulation, which incorporates the characteristic $(1 - p_t)^\gamma$ modulation to suppress well-classified examples:

\begin{equation}
\mathrm{FL}(p_t) = -\alpha_t (1 - p_t)^\gamma \log(p_t)
\label{e6}
\end{equation}

where $p_t$ represents the predicted probability for the ground truth class, $\alpha_t$ denotes the corresponding class weight, and $\gamma$ controls the focusing intensity. Through systematic evaluation, we establish $\gamma$ of 2.0 as optimal, providing sufficient emphasis on challenging cases while maintaining training stability. This configuration effectively balances sensitivity for malignant detection against overall classification performance.

Test-time augmentation enhances prediction reliability by averaging model outputs across systematically transformed test inputs~\cite{wang2018test}. Rather than applying random transformations, we employ a deterministic augmentation set tailored for lung nodule analysis through empirical validation.

Our transformation pipeline encompasses the original image alongside horizontal/vertical flips, 90-degree rotations, and modest brightness, contrast, and saturation adjustments of ten percent. These modifications capture natural variations in nodule presentation while preserving diagnostically critical morphological features. We deliberately avoid aggressive geometric distortions that could compromise shape characteristics essential for accurate diagnosis. The aggregation process combines predictions from all transformed versions using mean averaging in logit space:

\begin{equation}
    \mathrm{TTA}(x) = \frac{1}{N+1}\Bigl(f(x) + \sum_{i=1}^N f(T_i(x))\Bigr)
    \label{e1}
\end{equation}

where $x$ represents the input image, $f$ denotes the model function, $T_i$ corresponds to the $i$-th transformation, and $N$ indicates the total transformations applied. This logit-space averaging proves more robust than probability-based aggregation or voting schemes. Notably, MobileViT-XXS demonstrates the greatest sensitivity to augmentation, likely reflecting its transformer architecture's responsiveness to input variations. By applying TTA independently to each base model before ensemble fusion, we amplify both architectural diversity and augmentation-induced robustness.

\section{Training Setup}

We conducted all experiments on the Kaggle platform using an NVIDIA Tesla P100 GPU with 16\,GB VRAM, which provided sufficient computational power for training our complex ensemble architecture. Our implementation was built using PyTorch 2.5.1 with CUDA version 12.4 and Python 3.11.11, taking advantage of its dynamic computation graph capabilities that are well-suited for custom ensemble architectures.

After extensive experimentation and hyperparameter tuning, we established an optimal configuration for our Multi-Attention Stacked Ensemble (MASE) model. We used a batch size of 64 with input images resized to 96×96 pixels, training for a maximum of 200 epochs. To prevent overfitting, we implemented early stopping with a patience of 60 epochs, meaning training would halt if validation accuracy failed to improve for 60 consecutive epochs. We initialized training with a learning rate of 5×10$^{-4}$ and applied weight decay regularization at 1×10$^{-4}$ to improve generalization.

Our learning rate scheduling strategy employed cosine annealing warm restarts, which creates cyclical learning rate patterns that help the model escape local minima while maintaining overall convergence. The scheduler began with an initial period of 10 epochs, then doubled the cycle length after each restart (creating cycles of 10, 20, 40, 80, and 160 epochs), with a minimum learning rate floor of 1×10$^{-6}$. To work around GPU memory constraints while effectively increasing our batch size, we accumulated gradients over 2 steps before updating model parameters. For data augmentation, we applied MixUp with an alpha value of 0.4, which helps the model generalize better by creating synthetic training examples through linear interpolation between different samples.

\begin{table}[htbp]
\centering
\caption{Training Optimization Techniques}
\label{tab:optimization}
\begin{tabular}{|l|l|l|}
\hline
\textbf{Optimization Technique} & \textbf{Configuration} & \textbf{Purpose} \\
\hline
Optimizer & AdamW ($\beta_1=0.9$, $\beta_2=0.999$, $\epsilon=1 \times 10^{-8}$) & Decoupled weight decay regularization \\
Loss Function & Dynamic Focal Loss ($\gamma=2.0$) & Address class imbalance \\
Learning Rate Scheduler & CosineAnnealingWarmRestarts & Escape local minima \\
Gradient Accumulation & 2 steps & Effective batch size increase \\
Early Stopping & Patience = 60 epochs & Prevent overfitting \\
Data Augmentation & MixUp ($\alpha=0.4$) & Improve generalization \\
\hline
\end{tabular}
\end{table}

We selected AdamW as our primary optimizer following the approach of Loshchilov \& Hutter~\cite{loshchilov2017decoupled}, which extends traditional Adam optimization by implementing decoupled weight decay regularization. This choice was particularly important for our heterogeneous ensemble architecture, as AdamW better handles the varying scales of gradients across different network layers compared to standard Adam with L2 regularization. The optimizer parameters were set with beta values of 0.9 and 0.999 for the exponential decay rates, epsilon of 1×10$^{-8}$ for numerical stability, and the weight decay coefficient of 1×10$^{-4}$ applied in a decoupled manner.

For our loss function, we implemented Dynamic Focal Loss with a gamma value of 2.0 to address the class imbalance inherent in our lung nodule dataset. This loss function helps the model focus more attention on hard-to-classify examples while down-weighting the contribution of easy examples during training.

\begin{figure}[htbp]
    \centering
    \includegraphics[width=0.75\linewidth]{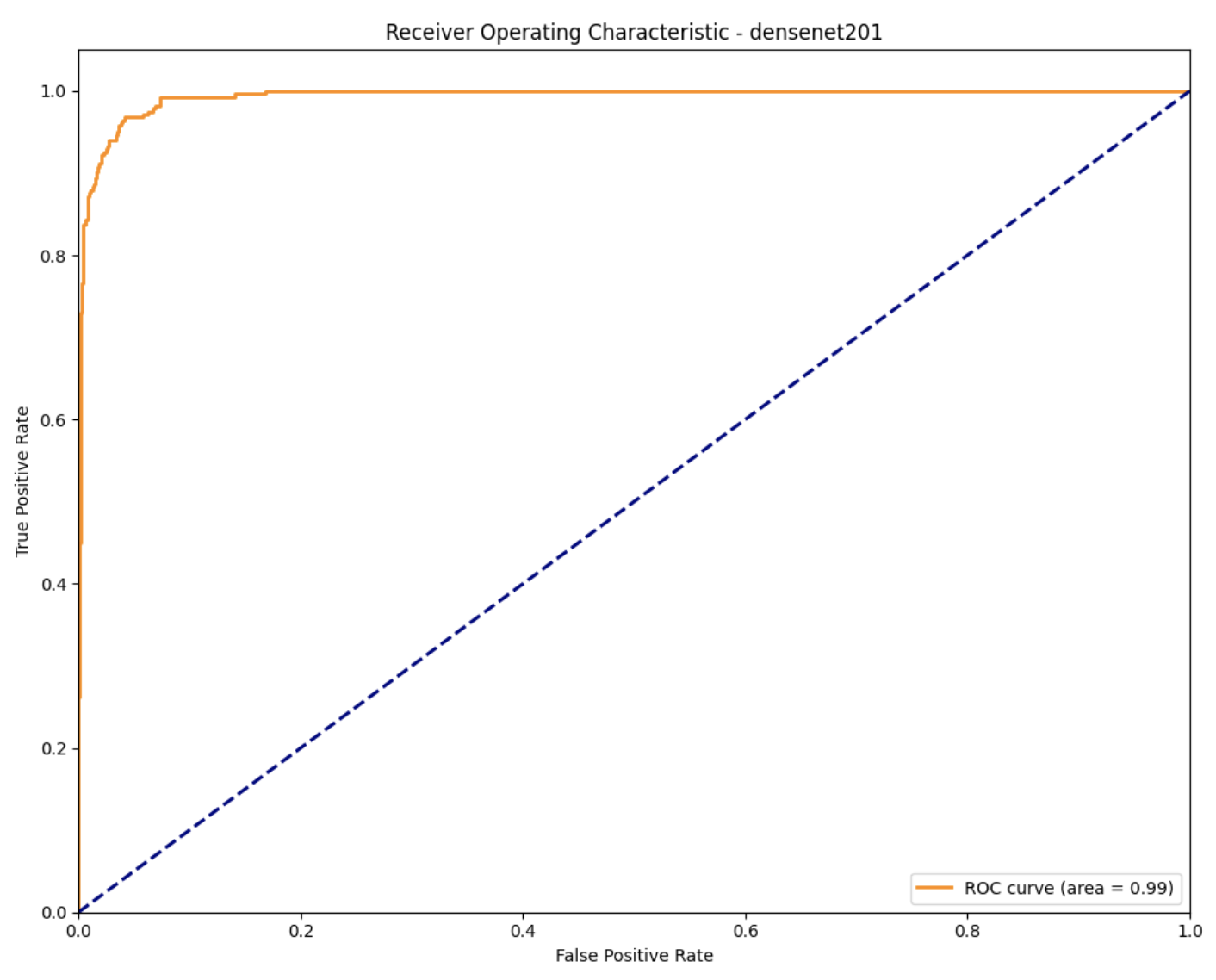}
    \caption{ROC Curve — DenseNet-201}
    \label{fig:roc_densenet}
\end{figure}

\begin{figure}[htbp]
    \centering
    \includegraphics[width=0.75\linewidth]{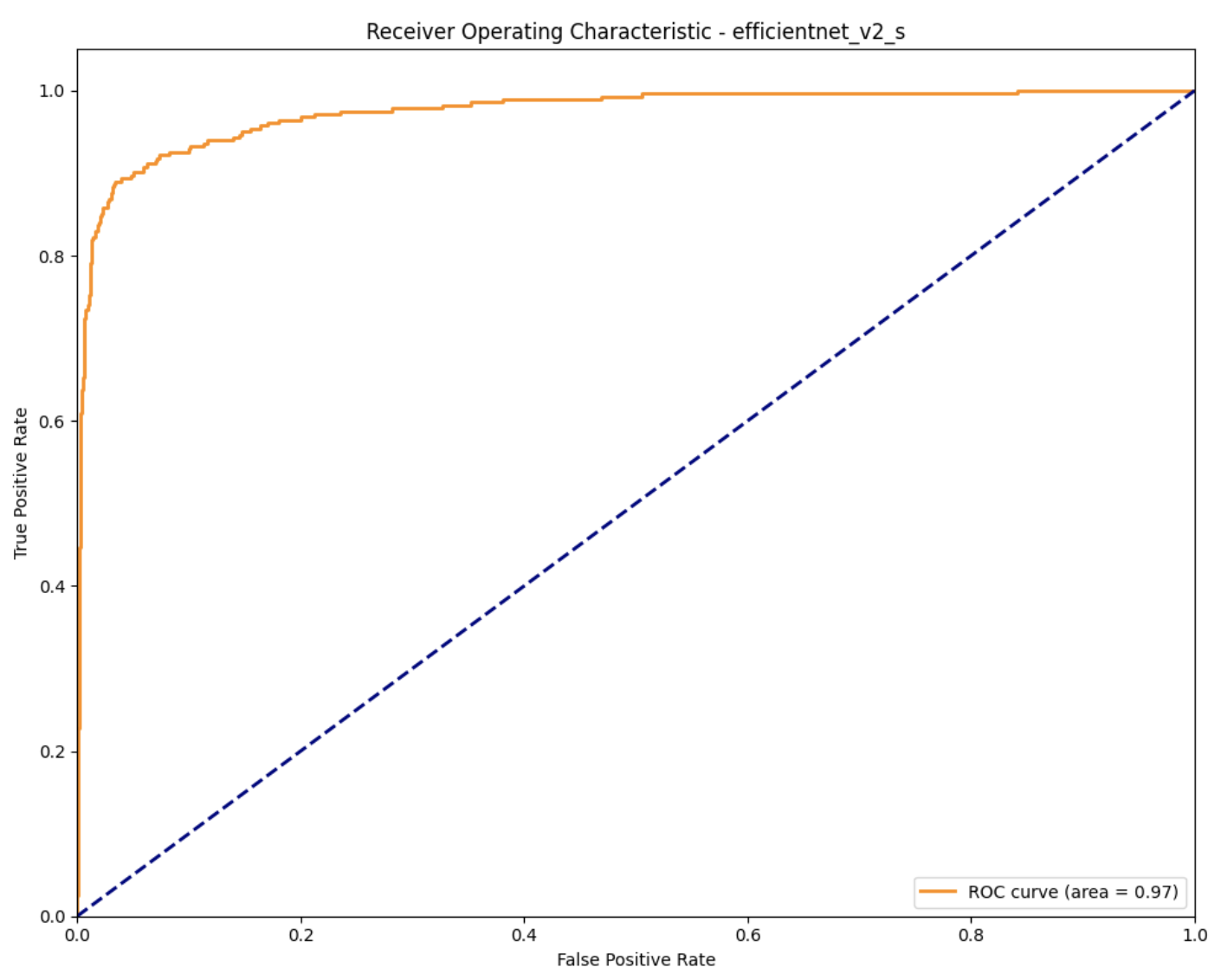}
    \caption{ROC Curve — EfficientNetV2-S}
    \label{fig:roc_efficientnet}
\end{figure}

\begin{figure}[htbp]
    \centering
    \includegraphics[width=0.75\linewidth]{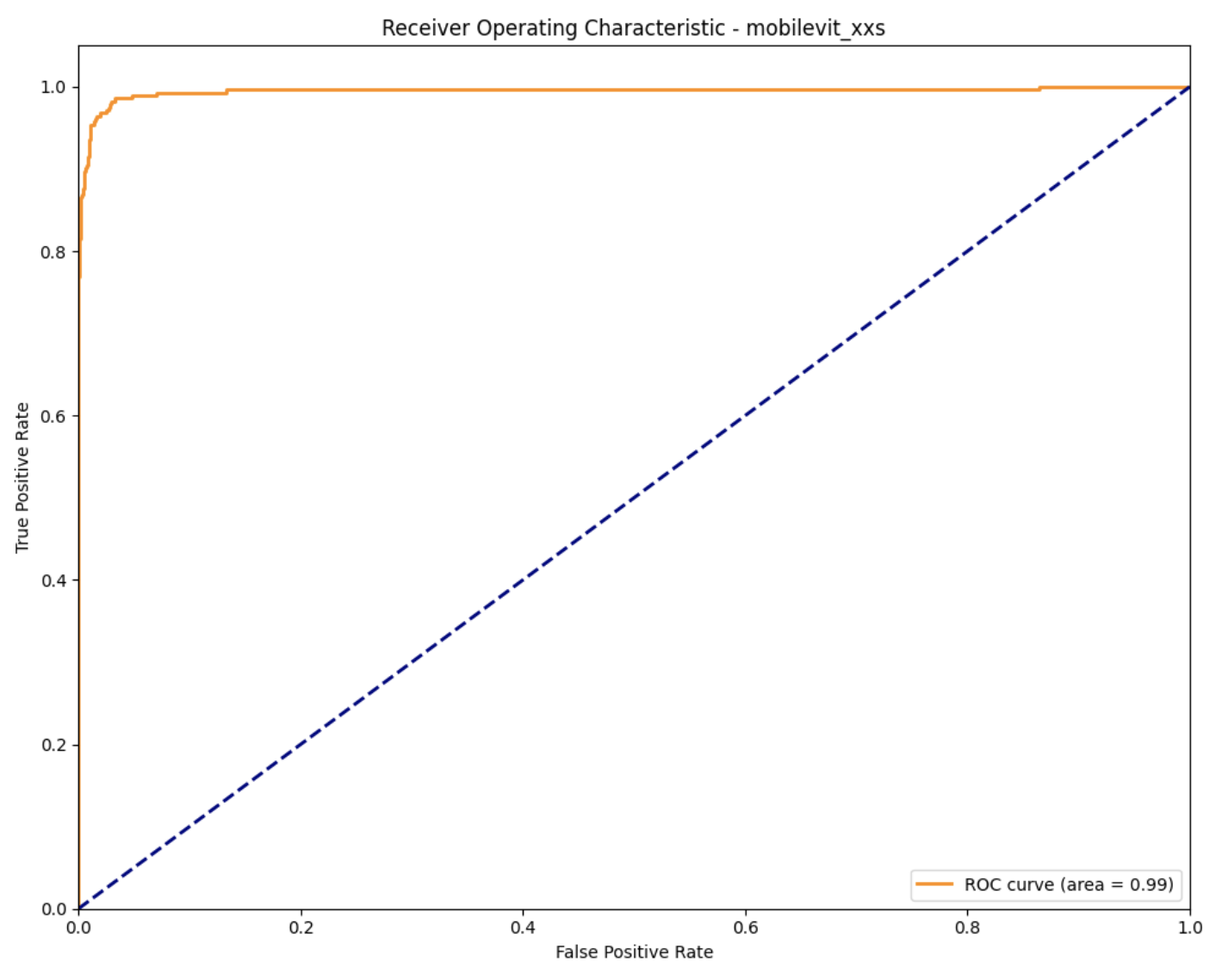}
    \caption{ROC Curve — MobileViT-XXS}
    \label{fig:roc_mobilevit}
\end{figure}

\begin{figure}[H]
    \centering
    \includegraphics[width=0.75\linewidth]{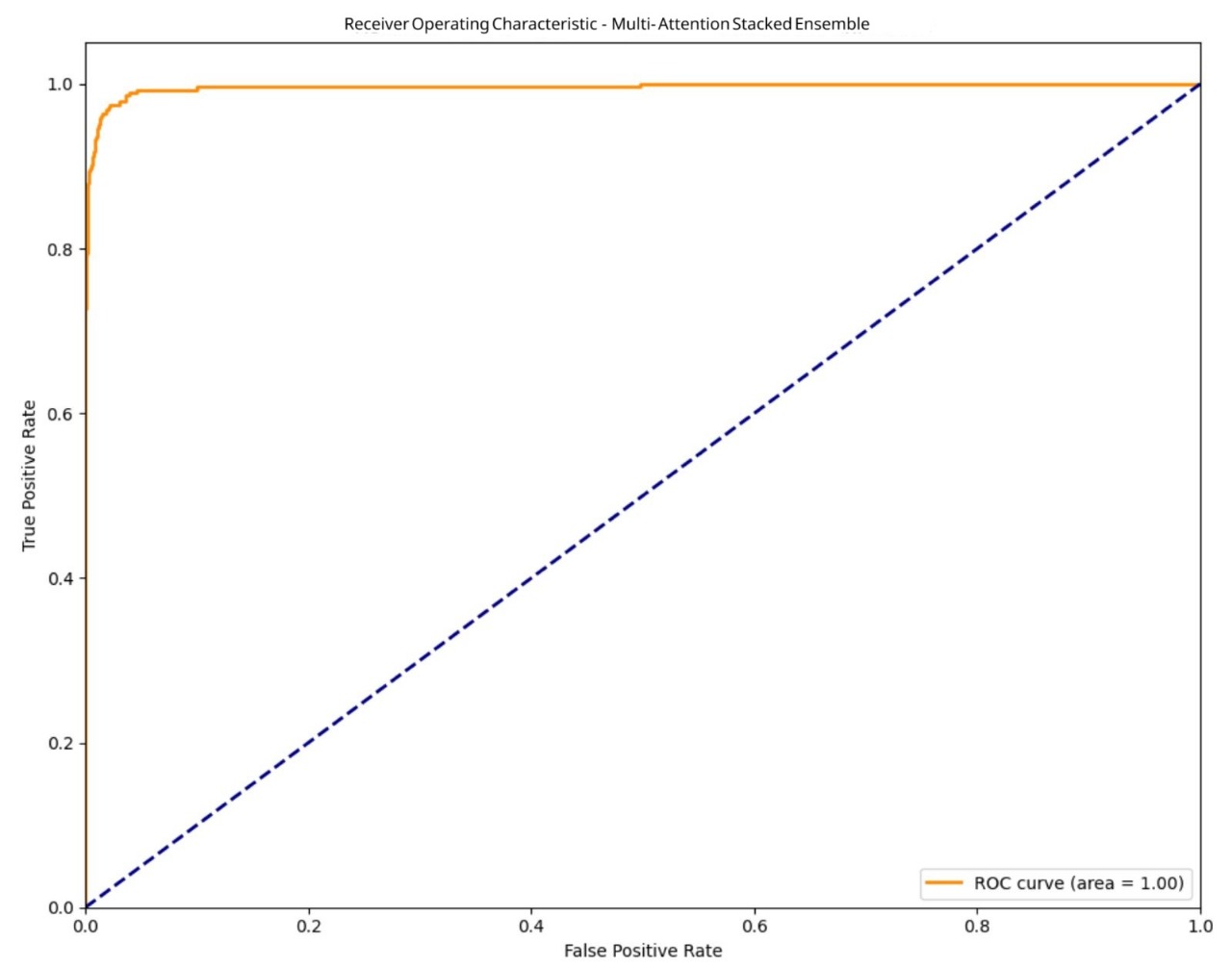}
    \caption{ROC Curve — Multi-Attention Stacked Ensemble (MASE)}
    \label{fig:roc_ensemble}
\end{figure}

For evaluation, we employed a comprehensive set of classification metrics to thoroughly assess our model's performance. We measured accuracy to understand overall correctness, precision and recall to evaluate class-specific performance, and weighted F1-score to account for class imbalance. ROC-AUC served as our primary metric for threshold-independent performance assessment, which is particularly valuable for medical applications where decision thresholds may vary based on clinical requirements. This metric also handles class imbalance well and provides a probabilistic interpretation of model performance.

\begin{figure}[h]
    \centering
    \includegraphics[width=0.8\linewidth]{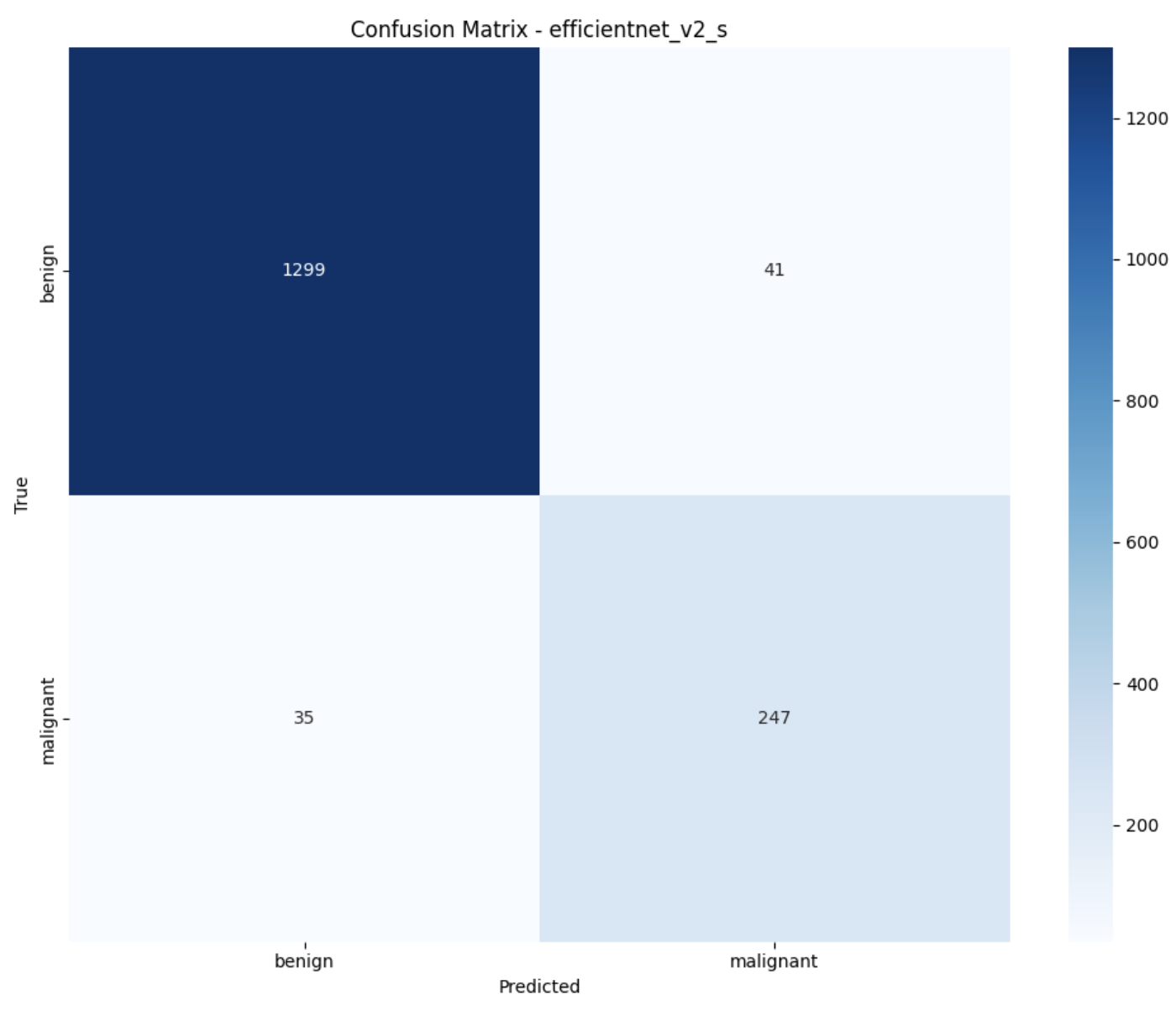}
    \caption{Confusion Matrix — EfficientNetV2-S}
    \label{fig:cm_efficientnet}
\end{figure}

\begin{figure}[h]
    \centering
    \includegraphics[width=0.8\linewidth]{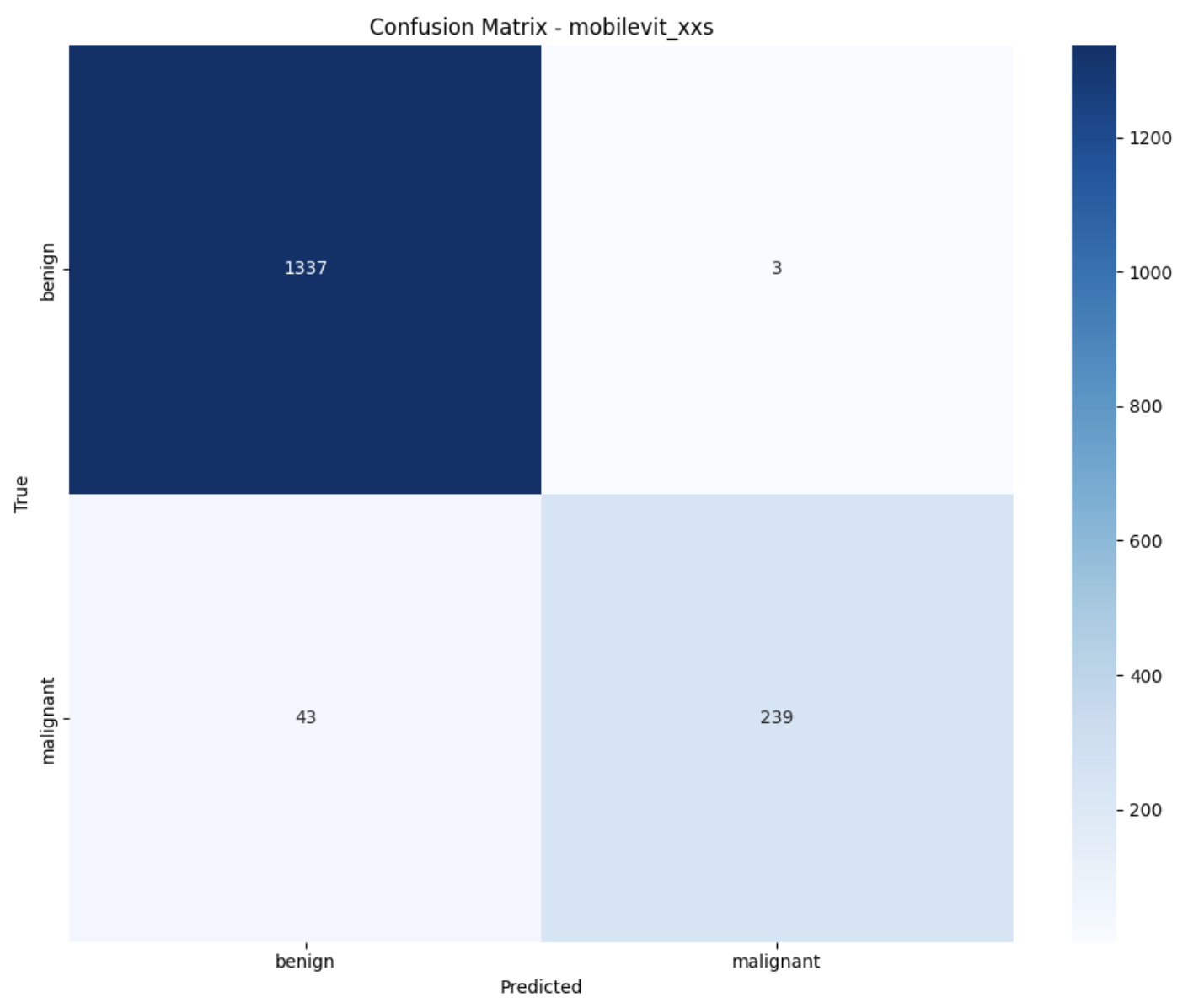}
    \caption{Confusion Matrix — MobileViT-XXS}
    \label{fig:cm_mobilevit}
\end{figure}

\begin{figure}[H]
    \centering
    \includegraphics[width=0.8\linewidth]{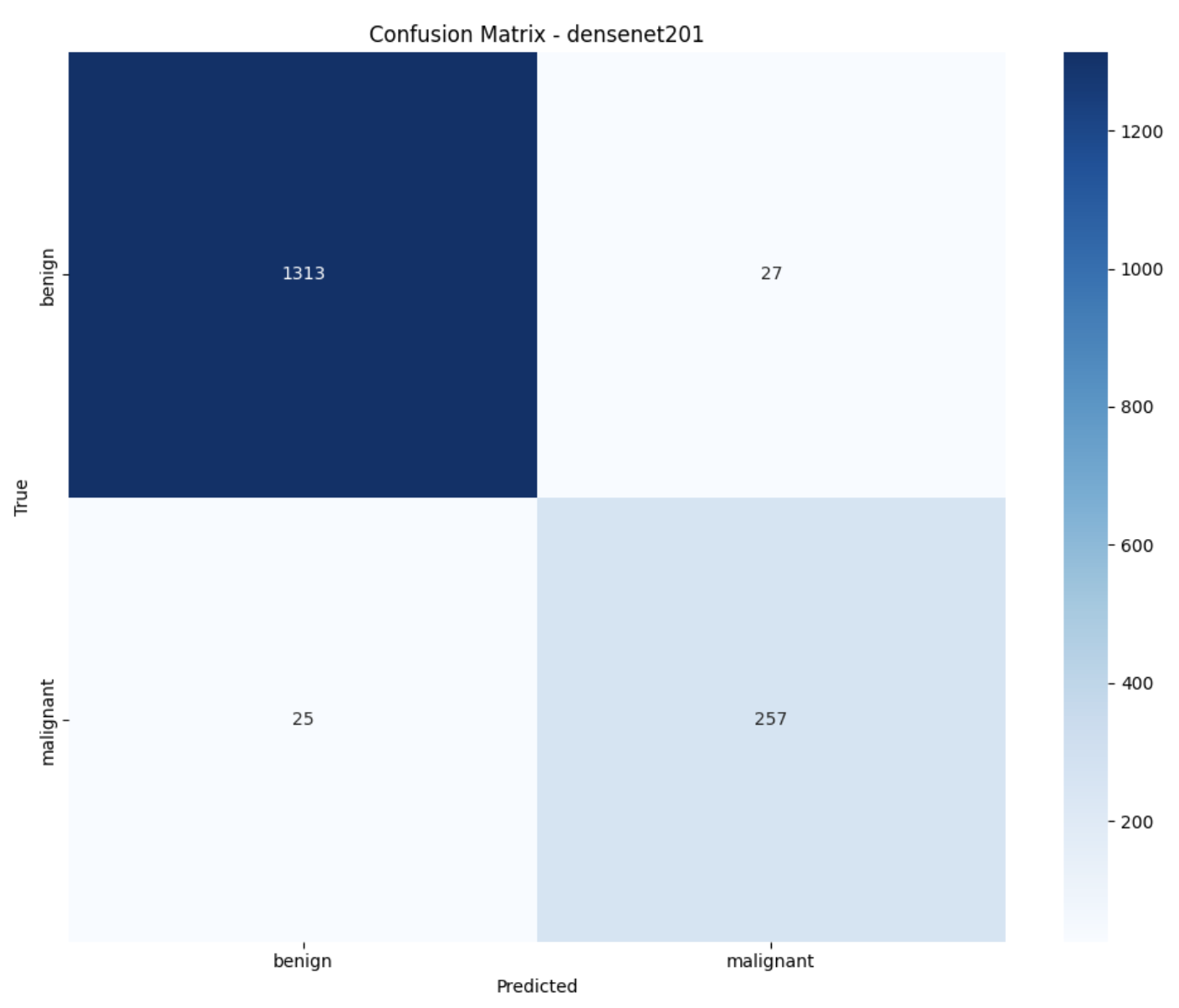}
    \caption{Confusion Matrix — DenseNet-201}
    \label{fig:cm_densenet}
\end{figure}

\begin{figure}[h]
    \centering
    \includegraphics[width=0.8\linewidth]{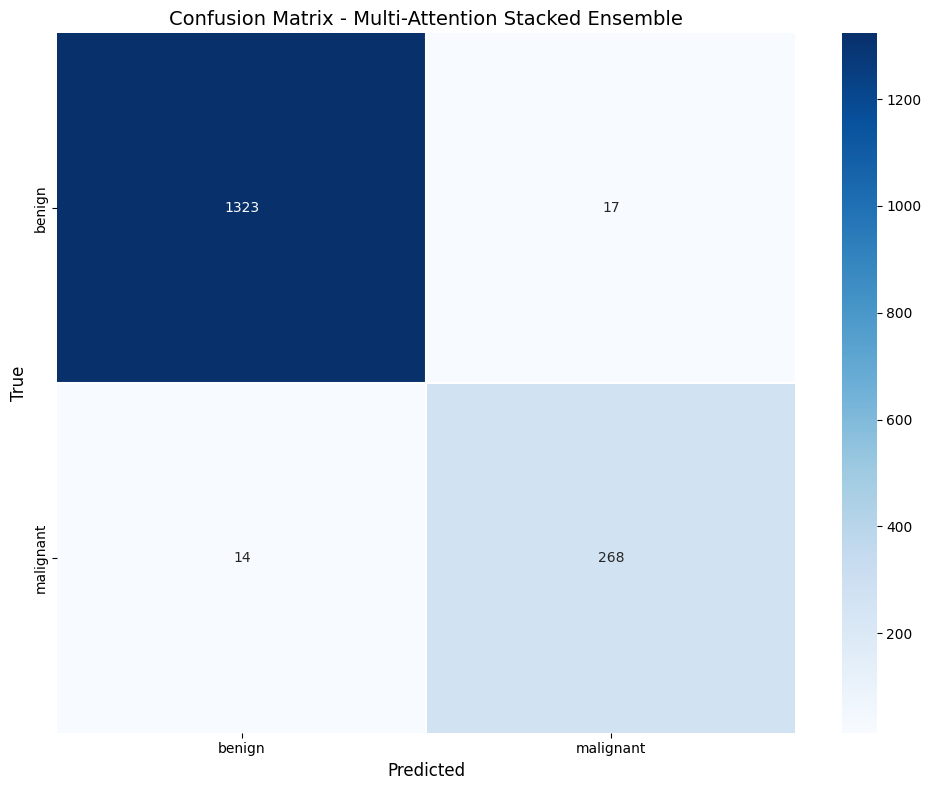}
    \caption{Confusion Matrix — Multi-Attention Stacked Ensemble (MASE)}
    \label{fig:cm_ensemble}
\end{figure}

We also generated detailed confusion matrices for each model to understand exactly where classification errors occurred. These matrices helped us identify whether models were more prone to false positives (incorrectly classifying benign nodules as malignant) or false negatives (missing actual malignant cases), which is crucial information for medical diagnostic applications where the cost of different types of errors varies significantly.

\section{Experimental Results}

This section presents the results of our lung nodule classification experiments using the Multi-Attention Stacked Ensemble (MASE) architecture, which achieved state-of-the-art performance with 98.09\% accuracy and 0.9961 AUC on the LIDC-IDRI dataset.

We evaluated three base models selected for their complementary architectures and feature extraction capabilities. Table~\ref{tab:individual_performance} summarizes their individual performance on the test set.

\begin{table}[h]
\centering
\caption{Individual model performance on LIDC-IDRI test set}
\label{tab:individual_performance}
\begin{tabular}{lccccccc}
\hline
Model & Accuracy (\%) & AUC & Precision (B/M) & Recall (B/M) & FP & FN & Total Errors \\
\hline
DenseNet-201 & 96.79 & 0.9936 & 0.98/0.90 & 0.98/0.91 & 25 & 27 & 52 \\
EfficientNetV2-S & 95.31 & 0.9744 & 0.97/0.86 & 0.97/0.88 & 35 & 41 & 76 \\
MobileViT-XXS & 97.16 & 0.9945 & 0.97/0.99 & 1.00/0.85 & 43 & 3 & 46 \\
\hline
\end{tabular}
\small B: Benign, M: Malignant, FP: False Positives, FN: False Negatives
\end{table}

DenseNet-201 achieved balanced performance through its dense connectivity pattern, which proved particularly effective at capturing subtle textural patterns like spiculation and irregular margins characteristic of malignant nodules. The model's feature reuse mechanism, where early layers' features are directly utilized by deeper layers, contributed to robust performance despite dataset imbalance, achieving nearly equal precision and recall for both classes.

EfficientNetV2-S achieved the lowest performance among the three models but still demonstrated strong results considering its parameter efficiency. The model excelled at detecting well-defined nodules with clear boundaries but showed limitations when dealing with ambiguous cases where subtle features determine malignancy. Its compound scaling approach enabled competitive performance particularly for benign nodules, though with a higher false negative rate that could be clinically concerning.

MobileViT-XXS emerged as the strongest individual model, leveraging its hybrid architecture that combines convolutional layers with transformer blocks. The transformer components enabled effective modeling of long-range dependencies within images, proving valuable for capturing relationships between nodule features and surrounding tissue context. However, the model exhibited a notable tendency toward high specificity, with 43 false positives but only 3 false negatives, suggesting a conservative approach to malignancy detection.

\begin{figure}[htbp]
    \centering
    \includegraphics[width=0.9\linewidth]{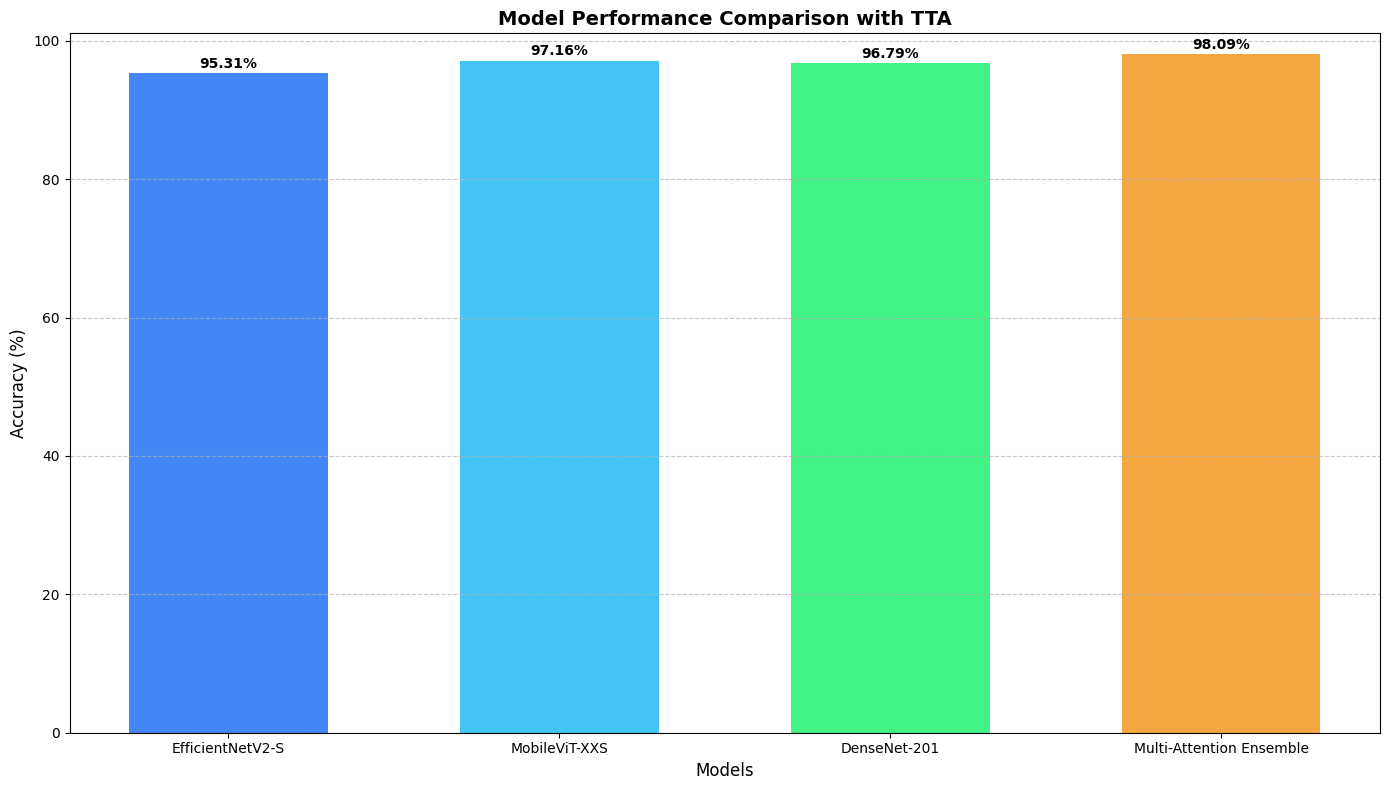}
    \caption{Accuracy comparison of individual models and the proposed Multi-Attention Stacked Ensemble (MASE) with Test-Time Augmentation (TTA). The MASE architecture outperforms all base models, achieving the highest accuracy of 98.09\%.}
    \label{fig:model_comparison}
\end{figure}

The proposed Multi-Attention Stacked Ensemble successfully integrated these complementary strengths to achieve superior performance. Figure~\ref{fig:model_comparison} illustrates the clear advantage of our ensemble approach over individual models.

Our ensemble achieved precision values of 0.99 for benign nodules and 0.94 for malignant nodules, with corresponding recall values of 0.99 and 0.95. The F1-scores of 0.99 and 0.95 for benign and malignant classes respectively indicate well-balanced performance crucial for clinical applications. The confusion matrix revealed only 31 misclassifications out of 1,622 test samples, with 17 false negatives and 14 false positives, representing a 35\% reduction in error rate compared to the best individual model.

The ensemble's error reduction was substantial across all base models: 32.6\% compared to MobileViT-XXS, 59.2\% compared to DenseNet-201, and 67.4\% compared to EfficientNetV2-S. While AUC improvements of 0.0016, 0.0025, and 0.0217 may appear modest, they represent significant gains in the high-performance regime where baseline AUC already exceeds 0.97.

To contextualize these results, we compared our approach against state-of-the-art methods from recent literature. Table~\ref{tab:sota_comparison} demonstrates MASE's superior performance across all evaluation metrics.

\begin{table}[!htbp]
\centering
\caption{Comparison of our proposed MASE with state-of-the-art approaches on LIDC-IDRI.}
\label{tab:sota_comparison}
\footnotesize
\setlength{\tabcolsep}{4pt}
\begin{tabular}{@{}llccccc@{}}
\toprule
Method & Architecture & Year & Accuracy & AUC & Sensitivity & Specificity \\
& & & (\%) & & (\%) & (\%) \\
\midrule
D.~Zhao et al.~\cite{sym10100519} & GAN + VGG16 & 2018 & 95.24 & 0.9800 & 98.67 & 92.47 \\
Xie et al.~\cite{xie2019semi} & Semi-sup. Adv. & 2021 & 95.68 & 0.9512 & 93.60 & 96.20 \\
A.~Halder et al.~\cite{Halder2022AdaptiveMorphology} & 2-Path CNN & 2022 & 95.17 & 0.9936 & 96.85 & 96.10 \\
Bushra et al.~\cite{Bushara2023LCDCapsule} & LCD CapsNet & 2023 & 94.00 & 0.9890 & 94.50 & 99.07 \\
Gautam et al.~\cite{gautam2024lung} & Weighted Ens. & 2024 & 97.23 & 0.9468 & 98.60 & 94.20 \\
\textbf{Proposed MASE} & \textbf{Multi-Att. Ens.} & 2025 & \textbf{98.09} & \textbf{0.9961} & \textbf{98.73} & \textbf{98.96} \\
\bottomrule
\end{tabular}
\end{table}

Our method achieved a 0.86\% accuracy improvement over the previous best results, with particularly notable gains in sensitivity. Recent approaches like Xie et al.'s semi-supervised adversarial model~\cite{xie2019semi} and Bushra et al.'s LCD-CapsNet~\cite{Bushara2023LCDCapsule} achieved strong results through architectural innovations, but our multi-attention integration strategy proved more effective. The dynamic attention mechanism adapts to each input, unlike traditional ensembles using fixed weights, while our dual-attention approach explicitly addresses both model-level and class-level weighting.

\subsection{Statistical Significance Test}

To validate statistical significance, we performed Wilcoxon signed-rank tests comparing our ensemble against each base model across 20 independent runs with different random initializations. The Wilcoxon test is a non-parametric statistical procedure that evaluates whether the median difference in paired observations equals zero~\cite{wilcoxon1992individual}, making it well-suited for model performance comparisons as it makes no assumptions about normality and is robust to outliers. We applied Bonferroni correction for multiple comparisons, adjusting the significance threshold to $\alpha = 0.0167$.

The ensemble achieved statistically significant improvements over all base models ($p < 0.001$ for all comparisons), demonstrating mean accuracy of 97.53\% ($\pm$0.41\%) versus 93.07\% ($\pm$4.13\%) for EfficientNetV2-S, 95.31\% ($\pm$4.07\%) for MobileViT-XXS, and 90.40\% ($\pm$8.61\%) for DenseNet-201. The most substantial improvement was against DenseNet-201, which showed highest variability ($\sigma = 8.61\%$). Even compared to the strongest individual performer (MobileViT-XXS at 95.31\%), our ensemble showed significant improvements ($p = 0.000006$). The Multi-Attention Stacked Ensemble achieved an average 4.60\% accuracy improvement with extremely low p-values ($p = 0.000004$ vs. EfficientNetV2-S, $p = 0.000006$ vs. MobileViT-XXS, $p = 0.000250$ vs. DenseNet-201), all well below the adjusted threshold, confirming robust and consistent performance gains.

The clinical implications of these performance gains are substantial. The 0.86\% accuracy improvement translates to approximately 8.6 fewer misclassifications per 1,000 nodules examined. Given millions of annual chest CT scans, this could prevent thousands of diagnostic errors~\cite{rubin2015lung}. Our enhanced sensitivity of 98.73\% for malignant nodules directly addresses the critical concern of missed cancers, where early detection can improve 5-year survival rates by 20-40\%~\cite{NLST2019ExtendedFollowUp}.

High specificity (98.96\%) reduces unnecessary follow-up procedures, each costing \$800-\$1,500 for additional imaging or over \$15,000 for invasive diagnostics~\cite{tanner2015management}.

\subsection{GradCAM++}

To better understand model decision-making, we employed GradCAM++ visualization, which produces high-resolution, class-discriminative explanations by analyzing gradient flow to the final convolutional layers~\cite{chattopadhay2018grad}. GradCAM++ was applied to each base model (DenseNet-201, EfficientNetV2-S, and MobileViT-XXS) to generate heatmaps highlighting the regions most influential in their predictions. Because, compared to the original GradCAM, GradCAM++ offers improved localization of multiple instances of the same class and generates more reliable explanations in complex medical imaging scenarios~\cite{kaur2022deep}. These visualizations reveal how each architecture focuses on distinct features of lung nodule images, offering valuable clinical insights into model decision-making and enhancing understanding of the Multi-Attention Stacked Ensemble's superior performance. Each heatmap is also accompanied by a confidence score based on the maximum softmax probability, reflecting the model's certainty in its prediction.

\begin{figure}[H]
    \centering
    \includegraphics[width=0.8\linewidth]{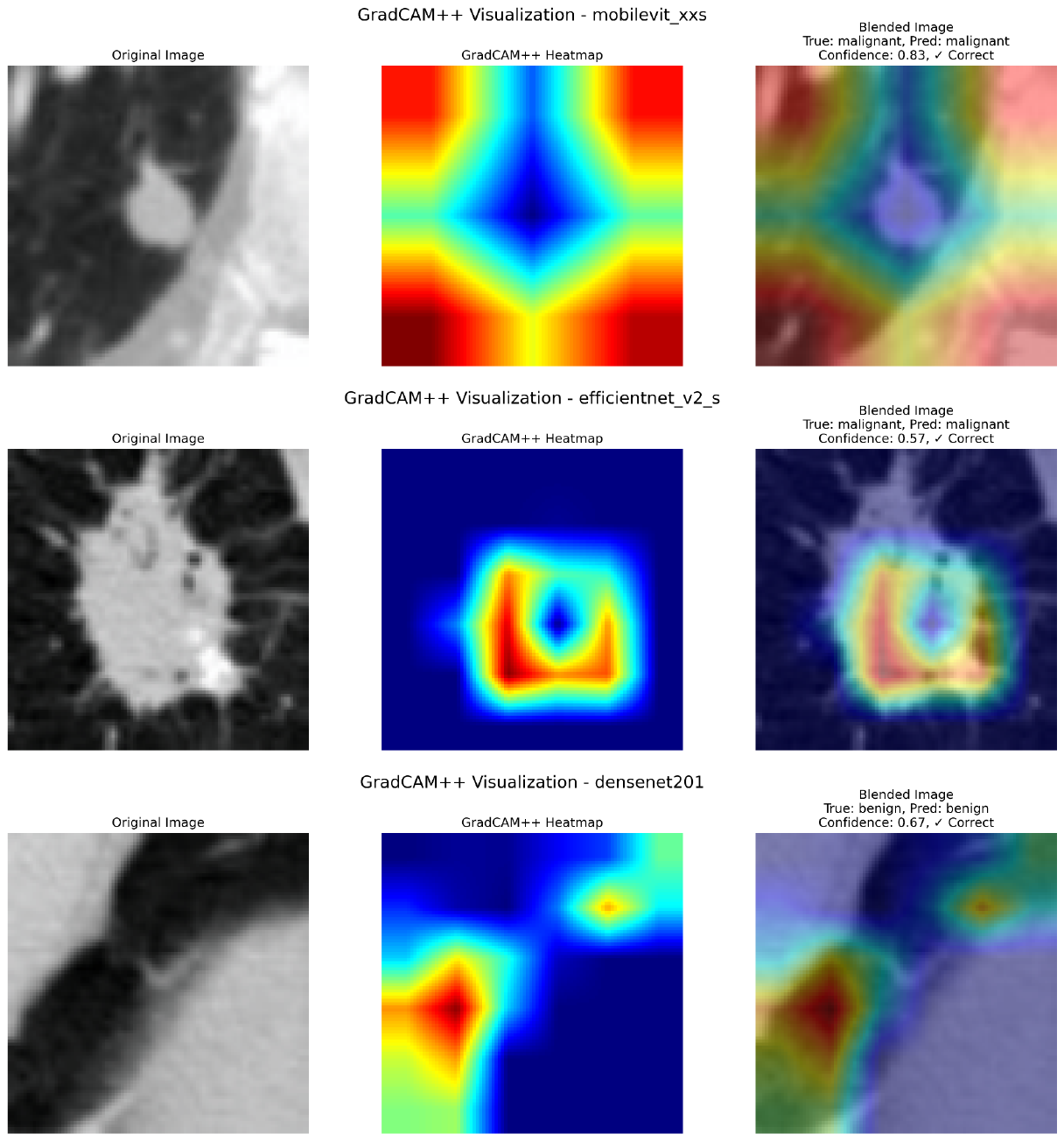}
    \caption{Comparison of GradCAM++ visualizations across base models for representative lung nodule samples.}
    \label{fig:gradcam_comparison}
\end{figure}

Figure~\ref{fig:gradcam_comparison} illustrates these activation maps, showing how the ensemble effectively combines the complementary feature extraction strengths of individual base models.

\section{Conclusion}

In this thesis, we presented the Multi-Attention Stacked Ensemble (MASE) for lung nodule classification, integrating model-level and class-level attention with a unified model adapter and Dynamic Focal Loss. On the LIDC-IDRI dataset, MASE achieved 98.09\% accuracy and an AUC of 0.9961, translating to a 35\% reduction in error rate over previous state-of-the-art methods while preserving high sensitivity (98.73\%) and specificity (98.96\%).

The key technical contributions include the dual-attention fusion mechanism that adaptively weights both model predictions and class logits, the custom 256-dimensional adapter head that ensures feature consistency and robust regularization. Together, these innovations enable the ensemble to dynamically emphasize the most reliable model or class output for each input, resulting in more nuanced decision boundaries and improved overall performance.

For future work, we envisage three main directions. First, extending MASE to multi-class Lung-RADS categorization would provide clinically actionable risk stratification aligned with reporting standards \cite{pinsky2015performance}. Second, we plan to experiment with alternative activation functions, such as the Gompertz function, in our attention and meta-learner modules, aiming to refine convergence behavior and calibration under class imbalance. Third, unifying nodule detection, segmentation, and classification into a single end-to-end architecture could leverage multi-task learning and enable extraction of quantitative biomarkers (e.g., volume, surface-to-volume ratio, texture heterogeneity) for richer clinical insights \cite{litjens2017survey}. Additionally, incorporating explainability techniques like feature attribution, concept-based explanations, counterfactuals, and uncertainty quantification will be essential for transparent decision support and regulatory acceptance in clinical workflows \cite{rudin2019stop}.

By combining advanced attention-based fusion, rigorous loss design, and explainability, MASE lays the groundwork for AI systems that not only match but complement radiologists’ expertise, moving closer to real-world deployment in lung cancer screening and beyond.

Our dynamic attention approach demonstrates a pathway toward adaptive AI systems that acknowledge medical data complexity. Despite implementation challenges including validation and regulatory requirements, this work provides a foundation for AI systems that complement human expertise in diagnostic processes, advancing toward collaborative healthcare AI that enhances radiologist capabilities.

These extensions will help translate MASE into robust, interpretable tools that augment radiologists’ expertise and advance AI‑assisted lung cancer screening.





\bibliography{sn-bibliography}

\end{document}